# Best Practices for Modelling Electrides

Lee A. Burton*[a]

Materials in which electrons occupy interstitial sites as anions are called electrides and exhibit unusual dimensionality-dependent electronic behavior. These properties make electrides attractive for catalysis, transparent conductors, and emergent quantum phenomena, yet their theoretical treatment remains challenging. In conventional materials, the ground-state atomic structure dictates the electronic configuration, whereas in electrides the electronic structure can instead govern the atomic arrangement. Here, the performance of commonly used exchange–correlation functionals is evaluated for representative one-, two-, and three-dimensional electrides. The results show that higher-cost approaches do not necessarily perform better across all cases, while standard methods capture the qualitative electride character and many key energetic and structural trends with surprising reliability. This behavior, likely arising from fortuitous error cancellation, supports the reliability of legacy studies in the field and the viability of efficient high-throughput exploration using low-cost methods. Overall, the findings support a tiered computational strategy for electride modelling, integrating system-specific heuristics with efficient first-principles screening. This approach balances computational feasibility with physical fidelity and underscores the continuing leadership of theory in the predictive discovery of electride materials across dimensionalities.

## Introduction

Electrides are materials in which electrons occupy lattice positions, acting as distinct anions rather than remaining bound to their parent nuclei. This electronic configuration imparts expected properties that are practically useful such as low work function and high ion mobilities,[1,2] which lend them to important catalytic applications such as ammonia synthesis,[3–5] $CO_2$ conversion and water splitting.[6] However they also exhibit properties that are less expected and more scientifically intriguing such as the unusual combination of metallicity and transparency in $Ca_{12}Al_{14}O_{32}$ (also called Mayenite),[7] off-atom void-centred magnetism in β-$Yb_5Sb_3$,[8,9] or off-atom delocalised magnetism in $Gd_2C$, [10] and surface free electron gas that persists even to the exfoliation of atomic-level layers.[11] These are all areas of active research, along with investigations about electride candidates for topological insulators,[12,13] and superconductors,[14] and so on.

The same unconventional behaviour that makes electrides scientifically fascinating *e.g.* electrons residing in interstitial space, exhibiting partial localization and strong correlation *etc*. also renders them especially difficult to describe.[15] The field of electride chemistry has long faced scepticism, owing to challenges in verifying both structural and electronic aspects experimentally. Arguably, the field as it pertains to inorganic materials reached maturity only recently with the direct experimental observation of anionic electrons. Palacios *et al.* claimed the distinction of "first experimental proof of their electride nature" referring to Mayenite in 2008.[16] However, these observations were based on model-dependent XRD Fourier reconstructions that revealed residual electron density at cage centres not a direct measurement of free electrons. The electronic-level confirmation came in 2016, when angle-resolved photo-emission spectroscopy (ARPES) performed on $Ca_2N$ resolved a dispersive, quasi-two-dimensional conduction band originating from interlayer electrons.[17] Prior to this milestone, the existence of "free" electrons in inorganic frameworks was inferred indirectly from neutron diffraction structural refinements that could show that cages, channels or layers were empty of atomic species that had been proposed on the basis of charge imbalance compensation.[18] Neutron diffraction was especially necessary for its ability to discern proton nuclei, as early electrides could be explained away as presupposed hydrides because the hydride ion was not trivial to observe in routine XRD.[19]

The paucity of experimental confirmation regarding the structure and properties of electride materials is a challenge that is unlikely to be resolved in the near future. Even if not intrinsically unstable, electrides are known to be highly sensitive to their environment.[5] Even brief exposure to moisture or oxygen can lead to rapid hydride formation or oxidation, resulting in irreversible degradation.[11,20] Consequently, the few electrides that exhibit genuine stability are especially valuable in contributing to the discourse. Notably, Mayenite remains the archetype of a stable electride even at high temperatures, pressure and extreme environments;[21] $Y_5Si_3$ has been reported to retain stability even after water exposure;[22] and $Hf_2S$ was recently reported to be water and acid stable.[23] It once seemed to be the case that the relative stability of electrides may correlate with the degree of exposure of the free electron density to the environment, which led to the convenient classification of considering electrides by their dimensionality.[24] Zero-dimensional (0D) electrides have anion electrons localised in cages, one-dimensional (1D) electrides have them delocalised in channels and two-dimensional (2D) electrides have them delocalised in planes. Fortunately, the three stable electrides mentioned correspond to each of these classifications respectively, not just injecting further enthusiasm to the field but affording theorists valuable benchmarks against which to test modelling methods.

[a] Department of Materials Science and Engineering, The Iby and Aladar Fleischman Faculty of Engineering, Tel Aviv University, Ramat Aviv, Tel Aviv 6997801, Israel.

Historically, modelling electride materials with standard electronic-structure methods posed theoretical challenges that paralleled experimental uncertainty. Electrides persistently defied heuristic wisdom that held for conventional materials such as charge neutrality, bonding paradigms, local coordination environments *etc*.[15] In the absence of a physically rigorous assignment of oxidation states or electronegativity,[25] theorists lacked the usual tools of chemical intuition on which classical electronic-structure models rely. Thus, the interpretation of electrides relies predominantly on first-principles electronic-structure methods, particularly density functional theory (DFT), which is not without its own limitations.

Usually in DFT, the atomic structure defines a ground-state electron density but with electrides the electron density informs the ground-state structure and this inter-dependent feedback-loop in turn defines the emergent properties for which electrides are sought after. As such, the usually relatively small effects of electron confinement, exchange–correlation and off-site magnetism can amplify to qualitatively erroneous results for electrides, which isn't the case for conventional materials. Commonly used exchange–correlation functionals in DFT such as the generalized gradient approximation (GGA) functional of Perdew, Burke, and Ernzerhof (PBE) are known to underestimate electron localization, [26,27] thus any predicted interstitial electron density must be interpreted cautiously, as it may reflect the functional's bias rather than a physically confined anionic electron. One of the standard approaches to mitigate this problem is with a Hubbard-correction, commonly known as DFT + U.[28] This scheme penalizes fractional orbital occupancies to restore localization but is inherently atom-centred and therefore ill-suited to describe diffuse or delocalized interstitial electrons. Higher order methods such as the so-called hybrid methods like HSE06 were empirically developed to correct for band gap values in semiconductors, [29–31] which is problematic given almost all electrides are expected to be metallic because of the free electron density and so can't be compared directly to experiment in this way (although, true-to-form, an exception exists in the semiconductor electride $Y_2C$).[32] Still, incorporating a fraction of exact exchange from the Hartree-Fock contribution generally yields what is expected to be a more reliable description of the electron density, albeit at significantly higher computational cost.[33] Treatments "higher" than hybrids such as self-consistent GW become prohibitively expensive and are reserved for a select few confirmations of already likely successes,[20] rather than being amenable to *e.g*. explorative screening. Furthermore, certain *ab-initio* software, such as the Vienna Ab Initio Simulation Package (VASP), achieves computational efficiency by imposing symmetry which is calculated from the atomic coordinates and may constrain the formation of the anionic electron if not deactivated.

All that is not to say that theoretical tools are not useful, quite the opposite. To the knowledge of the author, DFT has led directly to the experimental verification of at least four new electrides in the past- the first redox active electride: $Sr_3CrN_3$,[20,24] and the 2D electrides $Y_2C$,[34,35] $Hf_2S$[36,23] and $Y_2C$;[37,38] and will almost certainly do so again in the future (more examples are known for high-pressure electrides but they are considered to be outside of the scope of this work as non-ground state materials).[39,40] At the time of writing DFT studies have simulated thousands of candidate electrides,[15,41–45] and will always proceed at a rate far in excess of any possible experimental confirmation for the reasons already discussed. Fortunately, the field of DFT has also been undergoing its own development. Among the more recent advances is the meta-GGA class of functionals, which augment the electron density and its gradient with an explicit kinetic-energy density term to provide better property prediction than GGA.[46] A prominent example being the strongly constrained and appropriately normed (SCAN) functional, designed to satisfy known exact constraints on the exchange–correlation energy but sometimes numerically sensitive in practice.[47] Its successor, $r^2$SCAN, retains the same physics while reformulating the functional to improve numerical stability and convergence.[48] In doing so, $r^2$SCAN is thought to deliver more reliable energetics, structures, and electron-density localization than conventional GGAs at comparable cost, and has therefore been widely adopted in large-scale materials modelling. For example, the Materials Project database recalculated all of its entries with the $r^2$SCAN functional over the previous iteration of the database,[49] which was originally GGA (now referred to as the "next-gen" and legacy Materials Project respectively).

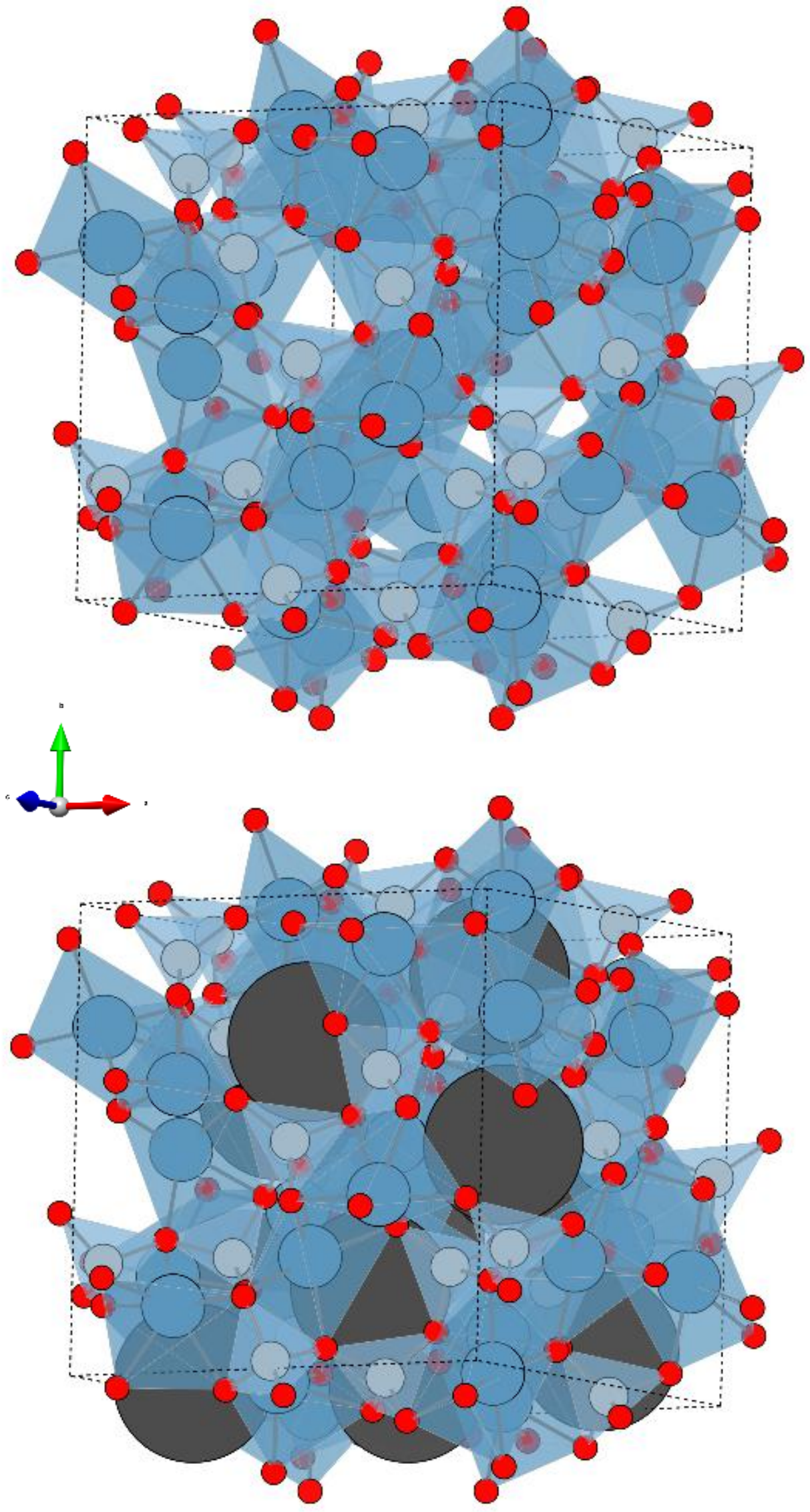

Figure 1: (Top) The conventional unit cell of Mayenite with unoccupied pore spaces. (Bottom) The same view of the conventional unit cell of Mayenite with all 12 pore spaces highlighted with artificial black spheres with radii of 5 Angstroms. In both cases, Ca ions are blue, Al ions are grey and O ions are red. The distorted Ca octahedra and Al tetrahedra can both be seen.

In this work, we evaluate how well different exchange–correlation functionals reproduce the experimentally observed structures and electronic characteristics of known electrides. Because electronic configuration and atomic structure are uniquely intertwined in electrides, agreement with experimentally observed relaxed structures serves as an efficient proxy for the underlying coupled structural and electronic physics. By benchmarking these functionals against experiment, the aim is to both identify which are most reliable for high-throughput screening of new electrides and to test whether electrides themselves provide a sensitive platform for probing the strengths and weaknesses of current electronic-structure methods. In this sense, electrides can serve not only as technologically promising materials but also as fertile ground for exploring new physics and refining the theoretical tools that describe it. To this end, six representative density functionals were tested:

- PBE – the GGA functional of Perdew, Burke, and Ernzerhof, serving as the baseline semi-local functional.[50]
- PBE-D3 – the PBE functional augmented with Grimme's D3 dispersion correction to account for long-range van der Waals interactions.[51]
- PBEsol – a revised GGA optimized for solids, intended to improve equilibrium volumes and lattice parameters.[52]
- $r^2$SCAN – the regularized form of the SCAN meta-GGA functional, designed to provide improved numerical stability and accuracy for diverse bonding environments.[48]
- $r^2$SCAN-rVV10 – the $r^2$SCAN functional combined with the rVV10 nonlocal correlation correction, incorporating dispersion effects within the meta-GGA framework.[53]
- HSE06 – the screened hybrid functional of Heyd, Scuseria, and Ernzerhof, which mixes a fraction of Hartree–Fock exchange with PBE correlation to improve band gap and electronic-structure predictions. [29–31]

Collectively, these functionals span the main rungs of "Jacob's ladder" of density functional theory,[54] allowing for the evaluation of both the reliability of current approximations for describing electride behaviour and the potential of electrides themselves as model systems for advancing exchange–correlation theory.

## Methods

All first-principles calculations were performed within the framework of density functional theory (DFT),[55,56] using the Vienna Ab initio Simulation Package (VASP).[57,58] The projector augmented-wave (PAW) method was employed to describe the core–valence interactions, and all structures were fully relaxed until the Hellmann–Feynman forces were below 0.005 eV Å$^{-1}$ on each atom and the total energy converged to within $10^{-6}$ eV.

For all calculations, plane-wave energy cutoffs of at least 520 eV were used, and Brillouin-zone integrations employed Γ-centred Monkhorst–Pack grids with a reciprocal-space resolution finer than 2π × 0.03 Å$^{-1}$. Ionic relaxations were performed for each functional to ensure internal consistency in comparing equilibrium geometries and electronic structures and were repeated iteratively until convergence was achieved within a single ionic step, ensuring the absence of Pulay stress and other numerical artifacts.[59] In the hybrid (HSE06) calculations, the screening parameter was set to 0.2 Å$^{-1}$ and 25% of exact exchange was included.[29–31] Spin polarization was enabled throughout, and all calculations were conducted without imposing symmetry constraints to allow possible electron localization associated with electride states.

Where applicable, long-range dispersion interactions were treated using either the semi-empirical Grimme D3 correction,[60] or the rVV10 non-local correlation functional,[53,61] depending on the underlying exchange–correlation approximation. The PBE-D3 method adds pairwise atom–atom dispersion terms to the GGA energy, providing an efficient account of van der Waals interactions, while the $r^2$SCAN-rVV10 scheme augments the meta-GGA $r^2$SCAN functional with a physically motivated non-local correlation kernel, ensuring a consistent and self-contained description of dispersion effects.

Table 1: the lattice parameters of ionically relaxed unit cell of $Ca_{12}Al_{14}O_{32}$ compared with the experimental values of Matsuishi *et al.*

| DFT method | Lattice parameter (Å) *a* | *b* | *c* | % difference with exp. *a* | *b* | *c* |
|---|---|---|---|---|---|---|
| PBE | 12.1024 | 12.1024 | 12.1024 | 0.87 | 0.87 | 0.87 |
| PBEsol | 11.9923 | 11.9923 | 11.9923 | -0.05 | -0.05 | -0.05 |
| PBE-D3 | 12.0234 | 12.0234 | 12.0234 | 0.21 | 0.21 | 0.21 |
| $R^2$SCAN | 12.0040 | 12.0040 | 12.0040 | 0.05 | 0.05 | 0.05 |
| $R^2$SCAN -rVV10 | 11.9772 | 11.9772 | 11.9772 | -0.18 | -0.18 | -0.18 |
| HSE06 | 12.0046 | 11.9938 | 12.0246 | 0.05 | -0.04 | 0.22 |

## Results

### 0D Electride Case

Mayenite is the archetypal electride, especially among the zero-dimensional cases. Chemically, Mayenite crystallizes as an ionic framework of Ca, Al, and O atoms in the *I-43d* space group (No. 220). This framework contains twelve cages per unit cell, each cage sharing its faces with four neighbouring cages to form a fully interconnected and permeable network, as shown in Figure 1. The cage diameter is approximately 5 Å, sufficient to accommodate a wide range of ionic species, although the narrow channels between adjacent cages can present bottlenecks for particularly large ions. The nominal composition of Mayenite is $Ca_{12}Al_{14}O_{33}$, but the framework can be more accurately described as a $[Ca_{24}Al_{28}O_{64}]^{4+}$ lattice charge-balanced by two $O^{2-}$ anions stochastically distributed among the twelve available cage sites.[62] When these oxide ions are removed, the excess electronic charge localizes within the cages, yielding the electride form of the material. [50] The *I-43d* symmetry is preserved only when the twelve pore sites are either all empty or equivalently occupied. The cage centres correspond to crystallographically equivalent points on a twelvefold special Wyckoff position. Partial occupation of these sites, where only one or a subset of cages is filled, necessarily lowers the overall crystal symmetry to a subgroup of *I-43d* symmetry.

Mayenite is notably stable with respect to chemical environment, pressure, and temperature, and has consequently

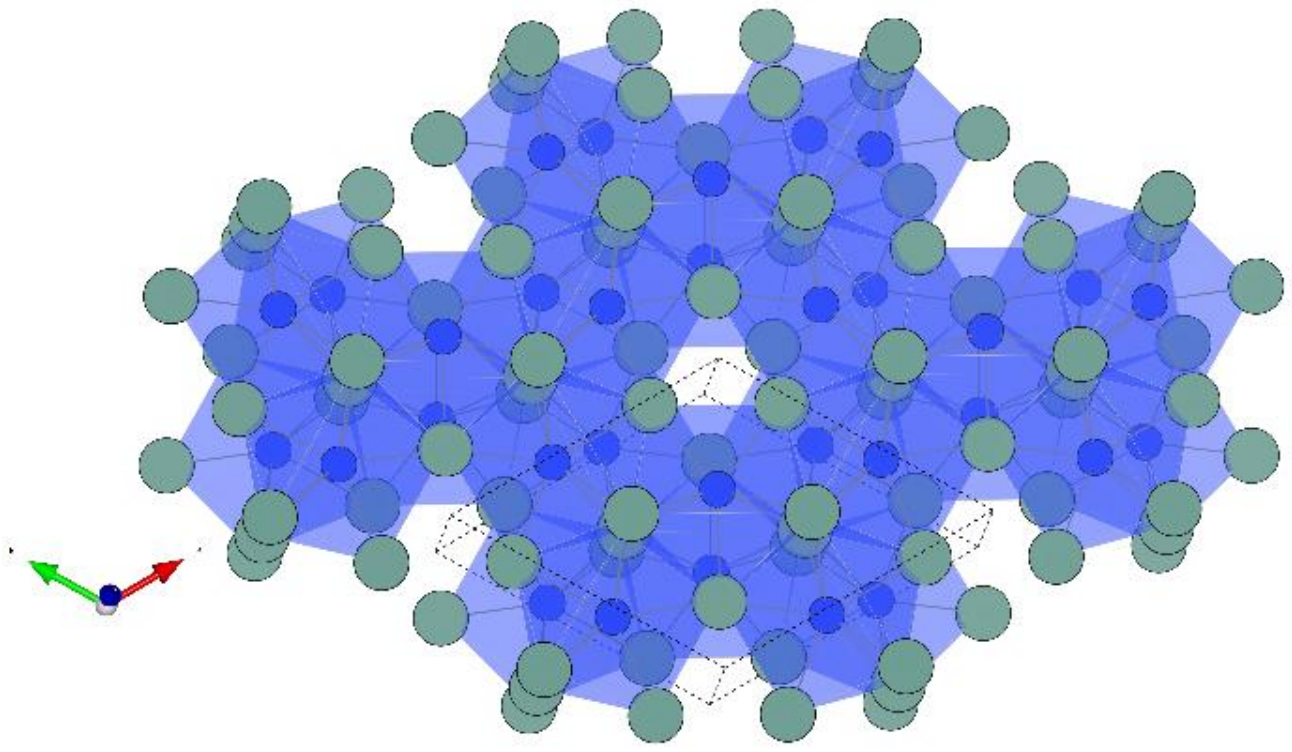

Figure 2: The unit cell of $Y_5Si_3$ with Y ions in green and Si ions in blue. The corners of the unit cell show the 1-dimensional void with Y ions coordinated to the anionic electron density

attracted attention as a robust support material for catalytic reactions, including ammonia synthesis.[51] Its combination of structural stability and extensive experimental characterization makes Mayenite an ideal benchmark system for assessing the accuracy of density-functional approximations applied to electrides. However, its relatively large unit cell poses a significant computational challenge for higher-order electronic-structure methods, which often become impractical except on state-of-the-art hardware.

In this work, we adopt the experimental lattice parameters reported from Rietveld refinement of combined powder X-ray and neutron diffraction data for the electride form $Ca_{12}Al_{14}O_{32}$, with unit cell dimensions $a = b = c = 11.9986$ Å and $\alpha = \beta = \gamma = 90°$ by Matsuishi *et al*.[63] These values agree closely with the earlier single-crystal X-ray diffraction measurements ($a = b = c = 11.97$ Å, $\alpha = \beta = \gamma = 90°$) reported in 1936 and with other literature reports.[62] Using this neutron-refined structure as a starting point, full structural relaxation is performed with the aforementioned exchange–correlation functionals to assess which best reproduces the balance between framework geometry and the active interstitial electron density that defines the electride state. The results of the relaxation and their comparison to experiment are shown in Table 1.

The relaxed lattice parameters for Mayenite show that most functionals perform well for this 0D electride, with deviations from experiment all below 1%. PBE exhibits the largest error, over-expanding the cell by ≈0.9% due to its well-known tendency to under-bind. In contrast, PBEsol and r²SCAN both reproduce the experimental lattice constant accurately (<0.05%), reflecting their improved treatment of exchange, correlation, and electron localization in a rigid oxide framework. The inclusion of dispersion corrections in PBE-D3 and r²SCAN-rVV10 leads to mild over-binding and small lattice contractions, though still around 0.2% of experiment. All relaxations were carried out without symmetry constraints; however, anisotropic lattice distortion emerges only for HSE06, consistent with the enhanced tendency of hybrid functionals to amplify small electronic asymmetries associated with the dilute anionic electrons in Mayenite. Overall however, the structural properties of Mayenite seem to be only weakly affected by functional-dependent electron delocalization.

Table 2: the lattice parameters of ionically relaxed unit cell of $Si_3Y_5$ compared with the experimental values of Roger *et al*.

| DFT method | Lattice parameter (Å) *a* | *b* | *c* | % difference with exp. *a* | *b* | *c* |
|---|---|---|---|---|---|---|
| PBE | 8.4517 | 8.4517 | 6.3751 | 0.50 | 0.50 | 0.49 |
| PBEsol | 8.3747 | 8.3747 | 6.2841 | -0.42 | -0.42 | -0.94 |
| PBE-D3 | 8.3701 | 8.3706 | 6.2763 | -0.47 | -0.46 | -1.06 |
| R²SCAN | 8.4571 | 8.4571 | 6.3955 | 0.56 | 0.56 | 0.82 |
| R²SCAN-rVV10 | 8.4420 | 8.4420 | 6.3697 | 0.39 | 0.39 | 0.41 |
| HSE06 | 8.4329 | 8.4329 | 6.3712 | 0.28 | 0.28 | 0.43 |

**1D Electride Case**

To represent the class of 1D electrides, yttrium silicide ($Y_5Si_3$) is selected as a compound experimentally established to host quasi-1D channels of interstitial electrons. Structurally, $Y_5Si_3$ crystallizes in a hexagonal $Mn_5Si_3$-type structure (space group $P6_3/mcm$, No. 193), composed of parallel channels running along the *c*-axis that form linear, electron-rich regions, as shown in Figure 2. These channels provide the confined yet extended topology characteristic of 1D electrides, where the interstitial electrons are delocalized along one crystallographic direction but spatially confined in the other two.

The crystal structure has been experimentally characterized in detail by Roger *et al.*[64] ($a = b = 8.4096$ Å, $c = 6.3437$ Å; $\alpha = \beta = 90°$, $\gamma = 120°$) whose findings agree closely with the subsequent neutron and X-ray diffraction study of Zheng *et al.*[65] ($a = 8.4087(1)$ Å, $c = 6.3422(1)$ Å $\alpha = \beta = 90°$, $\gamma = 120°$). These consistent results confirm the structural robustness of $Y_5Si_3$ and provide a reliable experimental foundation for theoretical benchmarking. From an electronic-structure perspective, the reliable comparison for a 1D electride is valuable because its interstitial electrons are hosted within well-defined, periodic channels rather than localized cavities as in Mayenite. This dimensional distinction provides a stringent test for exchange–correlation approximations: functionals that over-delocalize charge may smear the 1D electron density across the framework, while those that better balance exchange and correlation can capture the anisotropic confinement that defines the electride state. Here, the experimentally refined hexagonal $Y_5Si_3$ structure is subject to full structural relaxations using the same suite of exchange–correlation functionals as for the 0D Mayenite system. The direct comparison of each functional ground state and experiment is shown in Table 2.

The calculated lattice parameters for $Y_5Si_3$ are reproduced with good accuracy across most tested functionals, reflecting the intrinsic rigidity of the Y–Si framework. PBE and r²SCAN both slightly overestimate the lattice constants (≈0.5–0.8%), consistent with their mild under-binding, while PBEsol and especially PBE-D3 over-bind the structure, contracting all lattice directions; notably, PBE-D3 underestimates the c-axis by more than 1%, placing it well outside the accuracy achieved by the other methods. Among all functionals, HSE06 provides the best overall agreement with experiment, with deviations of only ≈0.3% in *a*,*b* and ≈0.4% in *c*, followed closely by r²SCAN-rVV10. Importantly, the impact of the interstitial anionic electrons can be seen in the *a*–*b* lattice parameters,

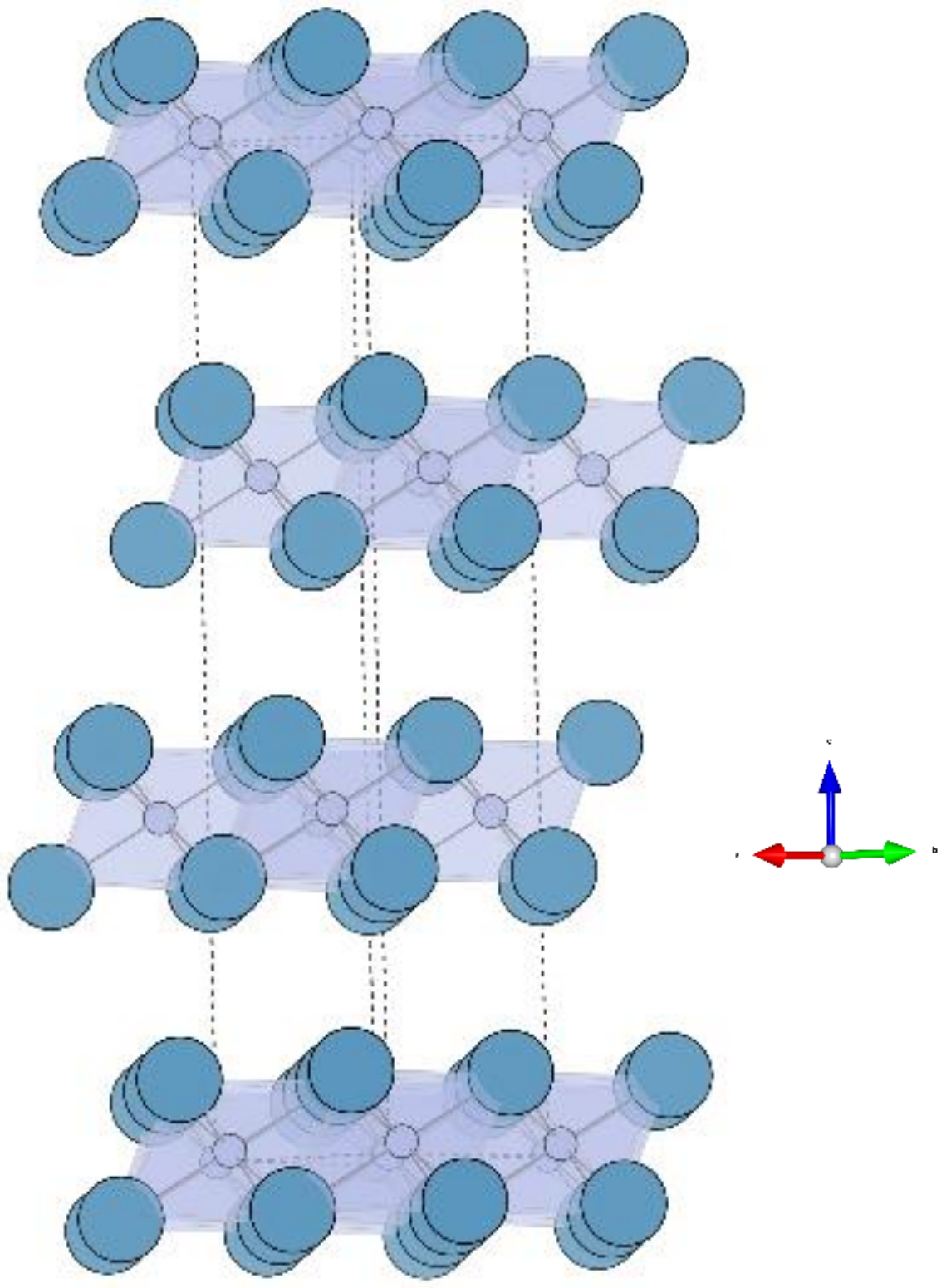

Figure 3: The unit cell of $Ca_2N$, showing Ca ions in blue and N ions in grey. The $[Ca_2N]^+$ layers separated by planes of interlayer electrons can be seen with N ions occupying the centres of octahedral coordination environments. These octahedra share edges to form extended two-dimensional layers

which define the transverse width of the quasi-1D channels. Functionals that over-localize electron density (PBEsol, PBE-D3) therefore compress *a* and *b*, artificially narrowing the channels, while those that over-delocalize (PBE, $r^2$SCAN) slightly expand them. The *c*-axis, along which the electrons are delocalized, is most sensitive with strong over-binding for PBEsol, PBE-D3 and strong under-binding for $r^2$SCAN similarly reflecting the balance between longitudinal electron pressure and Y-Si backbone bonding. The excellent overall performance of HSE06 and $r^2$SCAN-rVV10 indicates that exchange–correlation treatments which properly balance metallic bonding, nonlocal correlation, and self-interaction corrections most accurately reproduce both the structural framework and the anisotropic confinement of the interstitial electrons that define the 1D electride state. Notably, standard PBE seems to outperform $R^2$SCAN for this case, likely benefiting from a fortuitous cancellation of errors between its tendency to over-delocalize electrons and the intrinsic confinement provided by the 1D channel geometry.

**2D Electride Case**

For a 2D electride test case, $Ca_2N$ was chosen as one of the earliest and most thoroughly characterized layered electrides. While other layered electrides, specifically $Hf_2S$ mentioned in the Introduction, have been reported to exhibit exceptional thermodynamic and chemical stability, their discovery is comparatively recent and detailed structural, spectroscopic, and electronic data remain limited in the literature. By contrast, $Ca_2N$ benefits from decades of experimental and theoretical scrutiny, including comprehensive studies on its electronic structure, chemical reactivity, and exfoliation behaviour. [11,17] This extensive body of prior work makes $Ca_2N$ a more appropriate and reproducible choice for benchmarking density-functional approximations in 2D electride systems at this stage.

Structurally, $Ca_2N$ crystallizes in a rhombohedral $R\bar{3}m$ structure (space group No. 166) composed of alternating $[Ca_2N]^+$ cationic layers separated by interlayer planes that host the anionic electrons, as shown in Figure 3. These delocalized interlayer electrons form a quasi-two-dimensional electron gas confined between the calcium–nitrogen slabs, making $Ca_2N$ a model system for studying charge confinement and electron mobility in layered electrides. Note that in contrast to conventional two-dimensional materials, *e.g.* the popular layered transition metal dichalcogenide materials, the cations form the outer part of the ionically bonded layer.

The crystal structure of $Ca_2N$ has been experimentally characterized by Gregory *et al.* [66] and independently confirmed by Lee *et al.* [67] through single-crystal X-ray diffraction. The reported lattice parameters ($a = b = 3.6235$ Å, $c = 19.1015$ Å; $\alpha = \beta = 90°$, $\gamma = 120°$) are in excellent agreement with Lee *et al.*'s measurements ($a = b \approx 3.6$ Å, $c \approx 19.1$ Å; $\alpha = \beta = 90°$, $\gamma = 120°$), confirming the structural consistency across independent studies. These results establish $Ca_2N$ as a reliable experimental benchmark for computational validation. In this work, the experimental $Ca_2N$ structure is subject to structural relaxations using the same suite of exchange–correlation functionals applied to the 0D (Mayenite) and 1D ($Y_5Si_3$) cases.

The calculated lattice parameters for $Ca_2N$ are listed in Table 3 and compared with experiment, revealing systematic trends among the exchange–correlation functionals. In all cases, the in-plane a–b lattice parameters associated with the strong ionic–covalent bonding within the $[Ca_2N]^+$ layers are over-bound, yielding values that are consistently smaller than experiment. The persistence of this trend across semi-local, meta-GGA, dispersion-corrected, and hybrid functionals indicates that it is not a functional-specific artifact, but rather a general feature of the density-functional framework when applied to layered electrides. The absence of a similar effect in the 0D case suggests that the in-plane over-binding in $Ca_2N$ is tied to the presence of the extended interstitial electride electrons rather than to Ca–N bonding itself. Possibly, the approximate density functionals tend to over-stabilize Ca–N interactions by partially redistributing interstitial charge back onto the $[Ca_2N]^+$ layers, resulting in Ca–N bond lengths that are slightly shorter than observed experimentally. By contrast, more pronounced variations are observed in the out-of-plane lattice parameter *c*, which is predominantly governed by the much weaker interlayer interactions mediated by the confined anionic electrons and is therefore more sensitive to the treatment of exchange and correlation.

All methods reproduce the experimental in-plane lattice constant within 2%, which is relatively good but noticeably worse than the errors for other electrides in this study. Among all methods, PBE yields the closest overall agreement with experiment, slightly underestimating the in-plane lattice constants and modestly overestimating the interlayer spacing. In contrast, PBE-D3 and especially PBEsol both strongly over-bind, leading to a pronounced contraction of the lattice and significant underestimation of the interlayer distance. PBEsol therefore performs worst overall and should be avoided for modelling 2D electride systems. $r^2$SCAN follows the same trend as PBE, over-binding *a* and *b* and under-binding in *c*, but to a larger degree than PBE, resulting in a larger overall deviation despite its more advanced functional form. This systematic expansion likely reflects $r^2$SCAN's correction of PBE's over-delocalization, which reduces the unphysical spreading of the interlayer electron gas but simultaneously weakens interlayer cohesion. Adding non-local correlation in $r^2$SCAN-rVV10 partially compensates for this effect, greatly improving the description of the interlayer spacing with a modest increase in error for the bonded directions. HSE06 performs comparably to $r^2$SCAN in *a* and *b* but then under-binds in *c* the most of all functions tested. This behaviour reflects the reduced self-interaction error introduced by screened exact exchange, which weakens the attraction between the positively charged $[Ca_2N]^+$ slabs and the interlayer electride electrons. In addition, exact exchange enhances Pauli repulsion in the low-density interlayer region, further increasing the slab separation. Because HSE06 does not include explicit nonlocal correlation, these effects are not counterbalanced by dispersive attraction, resulting in a larger equilibrium *c*. The pronounced sensitivity of the interlayer spacing to the exchange treatment highlights the delicate nature of slab–electron interactions in two-dimensional electrides. These results indicate that PBE provides the numerically best match to experiment for $Ca_2N$. However, its success is likely coincidental, whereas $r^2$SCAN-rVV10 offers a more meaningful balanced and physically consistent picture of the weak interlayer binding and obtained agreement with experiment.

Table 3: the lattice parameters of ionically relaxed unit cell of $Ca_2N$ compared with the experimental values of Gregory *et al*.

| DFT method | Lattice parameter (Å) *a* | *b* | *c* | % difference with exp. *a* | *b* | *c* |
|---|---|---|---|---|---|---|
| PBE | 3.6105 | 3.6105 | 19.2288 | -0.36 | -0.36 | 0.67 |
| PBEsol | 3.5634 | 3.5634 | 19.0126 | -1.66 | -1.66 | -0.47 |
| PBE-D3 | 3.5692 | 3.5692 | 18.7469 | -1.50 | -1.50 | -1.86 |
| $R^2$SCAN | 3.6044 | 3.6044 | 19.3116 | -0.53 | -0.53 | 1.10 |
| $R^2$SCAN-rVV10 | 3.5934 | 3.5934 | 19.1943 | -0.83 | -0.83 | 0.49 |
| HSE06 | 3.6040 | 3.6040 | 19.3390 | -0.54 | -0.54 | 1.24 |

In practical terms, this distinction is important: functionals that over-localize or over-delocalize the interlayer electrons can respectively overestimate or underestimate exfoliation energies, charge confinement, and carrier mobility, which are outside of the scope of this study. The superior physical realism of $r^2$SCAN-rVV10 thus is still recommended to provide a firmer foundation for studying the more advanced properties for the 2D case.

## Discussions

### Comparison between 0D, 1D and 2D electride relaxations

Across the 0-, 1-and 2-dimensional cases, a coherent pattern emerges when comparing the ionically relaxed lattice parameters and those of experiment. Functionals that introduce empirical or semi-empirical corrections to strengthen equilibrium binding, such as PBEsol and PBE-D3, consistently over-bind and perform worst for the 1D and 2D electrides, where weak, non-local interactions are essential for reproducing the correct structural environment. In the 0D electride, where excess electrons are fully confined within isolated cages, all exchange–correlation functionals yield near-quantitative agreement with experiment, indicating that strong localization largely decouples the framework bonding from subtle electronic effects i.e. the excess electron does become more like a conventional anion. As dimensionality increases to the 1D case, the presence of extended electron channels introduces moderate anisotropy, and lattice parameters transverse to the channels become increasingly sensitive to how charge localization and partitioning are treated, with hybrid and dispersion-corrected approaches offering a more balanced description but only modest improvements over semi-local functionals. In the 2D electride, functional dependence becomes most pronounced: all tested functionals systematically over-bind the in-plane lattice parameters, while the interlayer spacing exhibits strong functional variability reflecting the delicate balance between slab–electron attraction, Pauli repulsion, and correlation effects. This monotonic increase in functional sensitivity from 0D to 2D, highlights an intrinsic limitation of current density-functional approximations in describing systems with increasingly delocalized interstitial electrons and underscores the importance of considering both physical fidelity and computational efficiency when modelling electride materials.

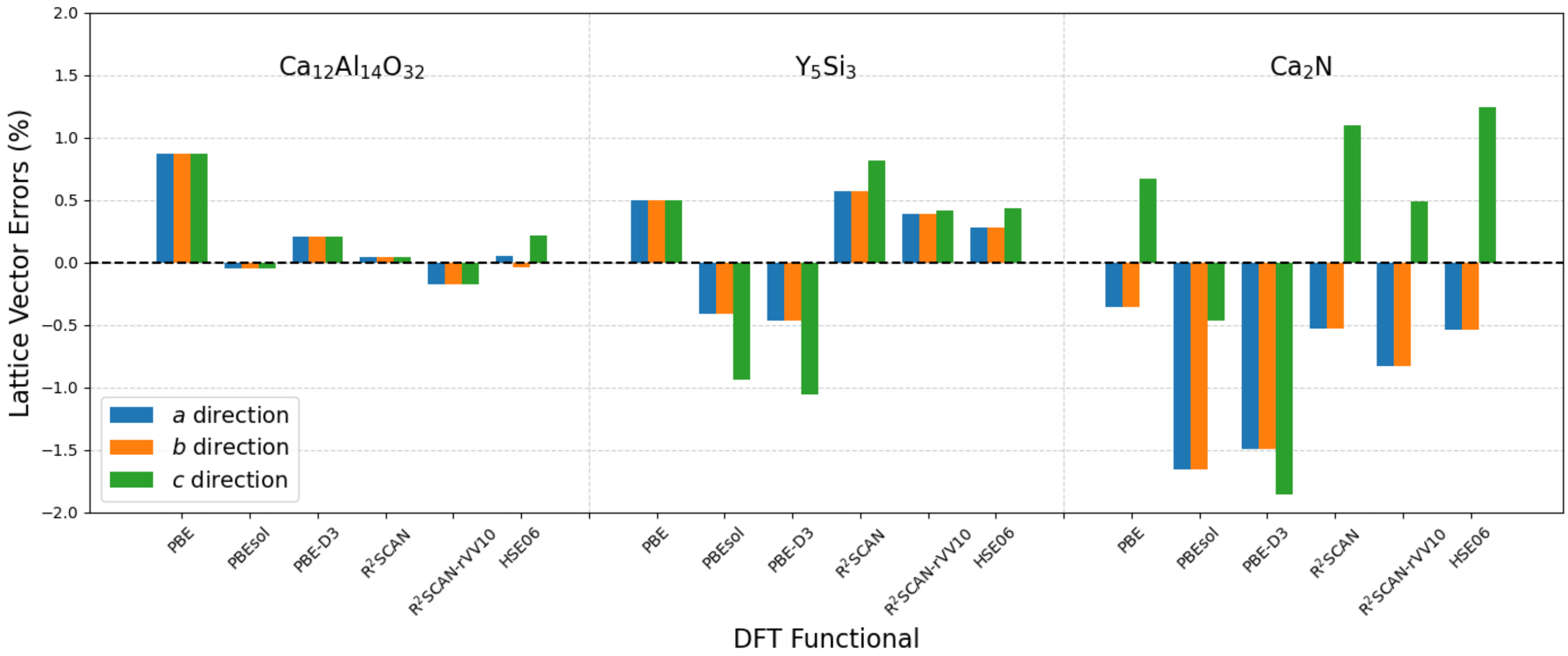

Figure 4: A graph showing the relative error values between converged ionically relaxed unit cell and experiment for each electride and every functional of this study. From left to right are the zero-, one- and two-dimensional electride cases respectively.

The cross-dimensional benchmarking allows for several observations. Firstly, HSE06, despite being the most computationally demanding method considered, does not consistently outperform lower-cost functionals, instead yielding the largest under-binding of the interlayer spacing in $Ca_2N$ and outperformed by other functionals for the remaining two electrides. Secondly, several of the functionals specifically optimised to work with relevant cases do not perform the best. To wit, PBEsol, specifically optimised for solids is not the best descriptor for Mayenite and PBE-D3, specifically optimised for layered materials, is not the best descriptor of $Ca_2N$. Finally, surprisingly good agreement obtained with PBE, which, despite likely arising from fortuitous cancellation of errors, nevertheless carries encouraging implications for the broader electride community. Much of the existing literature, where PBE has been the *de facto* standard, would appear to remain largely reliable, particularly for qualitative trends and structural predictions. Moreover, the ease of use, robustness, and low computational cost of PBE make it an attractive option for large-scale screenings, defect studies, and high-throughput materials discovery involving electrides. While more advanced functionals such as HSE06 and $r^2$SCAN-rVV10 provide a more physically faithful description across all dimensionalities, the fact that PBE often out-performs these functionals in terms of agreement with experiment means that computationally inexpensive workflows can continue to play a central role in electride modelling. In this sense, PBE's accessibility enables both broad exploration and continuity with prior work, even as higher-level methods refine understanding of the underlying electride physics.

**Identifying new electrides**

Electrides are physically defined by the presence of electrons that occupy interstitial regions rather than atomic orbitals, making real-space analysis essential for identifying and characterizing this off-atom electron density. While the materials studied in this work are well-established electrides, the assignment of any new electride materials continues to rely heavily on post-processing analyses of *ab initio* electronic-structure calculations. While many potential approaches exist,[68,69] two have most commonly been employed in the literature for this purpose: Bader charge analysis,[70,71] which partitions the total charge density, and the Electron Localization Function (ELF),[72] which highlights regions of electron localization. For all three materials considered in this study, both methods provide clear qualitative evidence for interstitial electrons even at the PBE level; however, consistent with recent work, neither yields a simple or universal quantitative figure of merit when applied in isolation.

The ELF provides a spatial map of electron localization and has long been used to visualize anionic electrons. For ELF, a number between zero and one represents how delocalised states are; zero being completely delocalised and one being completely localised. In electride materials, ELF isosurfaces can reveal compact regions of high localization that are spatially separated from atomic basins, corresponding to cavities, channels, or interlayer regions. These features can be examined using planar slices through the crystal lattice with modern visualization tools such as VESTA,[73] allowing apparently empty regions of the structure to be interrogated directly, or simply visualised as an isosurface in 3-dimensions, see Figure 5a. In general, improved exchange–correlation treatments lead to sharper and more localized ELF maxima. However, ELF values do not correspond directly to electron density and cannot be integrated to yield electron counts or oxidation states. ELF is therefore best understood as a qualitative, topological descriptor of chemically meaningful features rather than a quantitative measure of charge.

Bader charge analysis, by contrast, partitions the total electron density into zero-flux basins and can, in principle, yield quantitative charges through integration of the charge density. When a sufficiently fine fast Fourier transform (FFT) grid is employed, interstitial electrons may appear as distinct non-nuclear basins, and the integrated charge of such basins can serve as a quantitative indicator of electride character. However, as discussed in a recent work by Weaver *et al.,*[74] conventional Bader analysis applied to the total charge density often fails to identify electride electrons as separate basins, particularly when the interstitial charge density is small or spatially complex and is dominated by nearby atomic contributions. This limitation is especially evident in $Ca_2N$, which has a formal oxidation state of +1: in the default Bader partitioning, only the nine atomic basins (six Ca and three N per unit cell) are identified, with no explicit non-nuclear basin reported and total charge conservation maintained. Importantly however, this does not imply the absence of electride electrons in the charge density itself. By explicitly examining the topology of the charge density, non-nuclear critical points associated with interstitial electrons can be directly identified using the same method, even for $Ca_2N$, as shown in Figure 5b.  Thus, while standard basin-based Bader partitioning may fail to assign a separate interstitial basin, the electride electrons are nonetheless explicitly observable in the charge-density topology as critical points. These are the coordinates used in previous high-throughput screening studies that successfully identified new electrides.[15,24]

Weaver *et al*. subsequently propose the BadELF algorithm as an alternative strategy to address the challenges associated with assigning charge to interstitial electrons,[74] primarily by modifying how integration volumes are defined. In conventional Bader

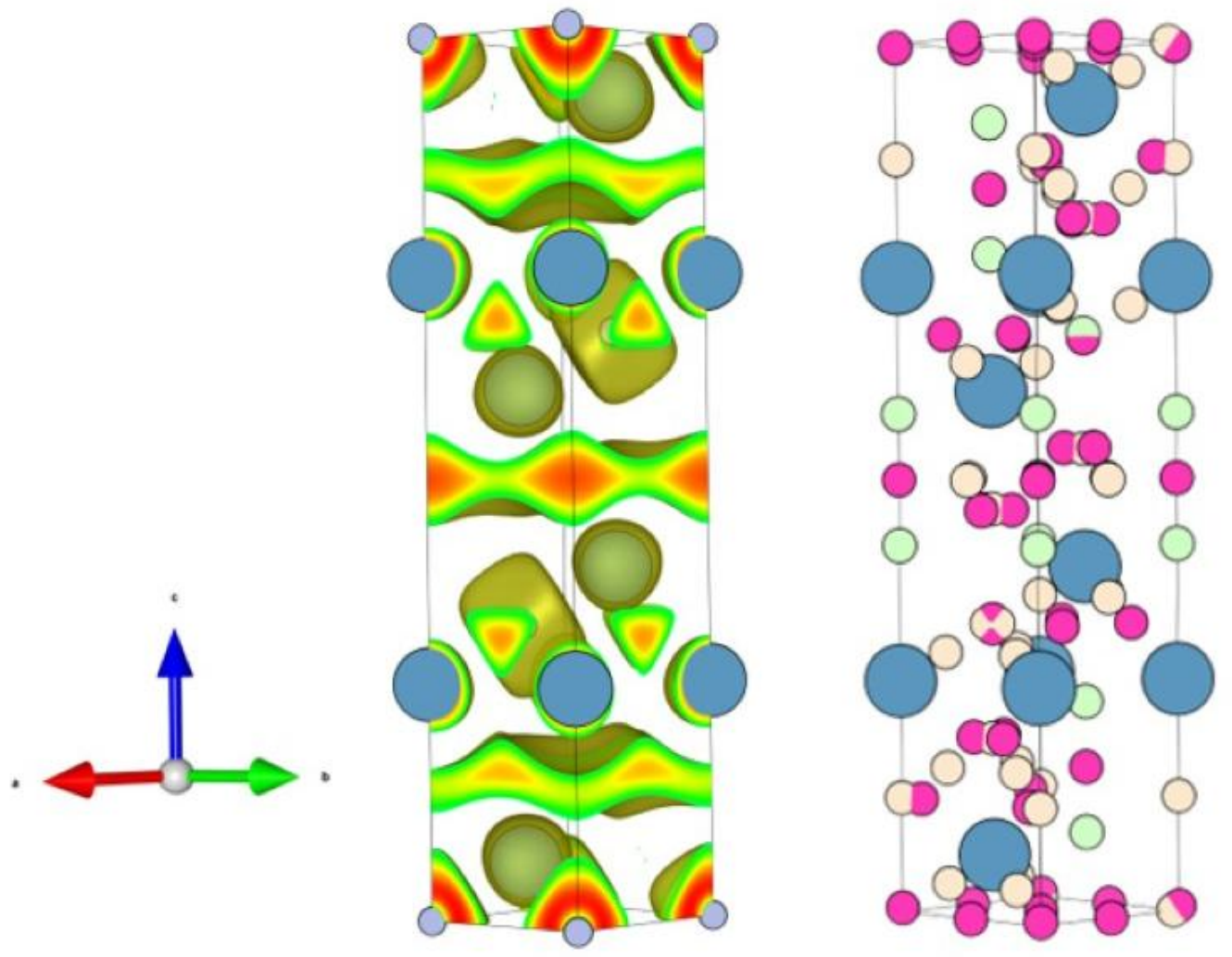

Figure 5a (left) Comparison of the electron localisation function (left) and Bader charge analysis and 5b (right) outputs when performed on the material $Ca_2N$ simulated at the PBE level. Each show the undulating interlayer electron density in different ways.

analysis, zero-flux surfaces are constructed directly from the total charge density, whereas the BadELF method instead defines atomic and interstitial regions using zero-flux surfaces of the electron localization function and subsequently integrates the charge density within these ELF-derived volumes. In this sense, BadELF represents an ELF-partition–driven charge integration scheme, while standard Bader analysis is a charge-density–partition–driven scheme. Both approaches integrate the same underlying charge density but emphasize different aspects of the electronic structure, localization topology versus charge-density topology, and can therefore yield different numerical interpretations when applied to electrides. Importantly, as shown in the present work, many of the limitations attributed to conventional Bader analysis arise from its standard basin assignment rather than from an absence of electride features in the charge density itself and these can be recovered either way.

Both the electron localization function (ELF) and Bader charge density analysis can provide strong evidence for electride character, but neither alone is definitive. Particularly for metallic systems, heuristic judgement remains necessary to distinguish electrides from conventional materials. Their combined use therefore provides a robust and internally consistent picture of electride behaviour, with ELF revealing the spatial localization and dimensionality of interstitial electrons and Bader analysis constraining charge partitioning between atomic and interstitial regions. The ELF visualizations and Bader charge-density analysis for the $Ca_2N$ at the PBE level from this study are shown in Figures 5a and 5b.

Finally, in pseudopotential-based calculations, the frozen-core approximation may become more consequential when electrons are not tightly bound to atoms. For example, standard PAW datasets for aluminium in VASP (the POTCAR file) treat only three valence electrons explicitly, a choice that is well justified for conventional bonding environments but may warrant additional scrutiny in electride systems hosting interstitial electrons, e.g. Mayenite. Again, no definitive rule can be prescribed, such assessments ultimately relies on the informed judgement of the materials scientist.

**Electrides as a test-bed for materials science**

The value of electrides as a stringent test bed for both established and emerging density functionals is underscored by several trends observed in this study. Notably, PBEsol performs among the worst for the one- and two-dimensional electrides, despite being explicitly optimized for solids—a category to which these materials formally belong. Moreover, the spread of relative errors increases with electride dimensionality, with both positive and negative deviations becoming more pronounced in higher-dimensional cases. In the specific case of the two-dimensional electride, PBE-D3 performs particularly poorly, highlighting how functionals designed for ostensibly applicable systems, such as layered materials, can fail most severely when confronted with the unconventional bonding environments characteristic of electrides

It might seem that electrides could simply be sidelined in this regard as outlier materials. However, recent attempts to distinguish electrides from ordinary materials have repeatedly failed along several fundamentally different pathways. Firstly, electrides and more conventional materials have been shown to be indistinguishable on the basis of off-site electron density, see *e.g.* materials with stereo-active lonepairs;[75] with one study going so far as to say that 3D electrides and intermetallic electrides can't be distinguished from ordinary metals at all.[5] Secondly, the Farbe centre (F-centre) defect, commonly observed in ionic solids, can be viewed as the dilute limit of electride behavior, with the distinction between a defective material and an electride corresponding to an arbitrary concentration of such defects.[76,77] Thirdly, electron deficient electrides are known,[78] thus it can't be that they are simply electron-rich materials as is sometimes stated in the literature.[74] Likewise, neutral electrides have been reported,[79,80] demonstrating that electrides cannot be generally described as over-valent, despite this characterization also frequently appearing in the literature.[15] Finally, analysis has shown that all groups of elements, are represented among electrides; not just alkali metals and not just lanthanides and heavy elements, but d-block metals, halogens and chalcogenides *etc* are represented in electride datasets.[15] Certainly, the elements in this study are by themselves not rare or understudied (Ca, O, Al, Y, Si and N) and they should all have their physics well accounted for with established methods. Either way, any attempt to dismiss the physics of electrides as not relevant to materials science in the broader context should be resisted and method developers are encouraged to capitalise on this opportunity to learn more about the fundamental physics that leads to the emergence of anionic electrons in the solid state.

## Conclusions

In this work, a systematic, cross-dimensional assessment of exchange–correlation functionals for modelling electrides spanning 0D, 1D, and 2D electronic topologies is performed. By comparing structural relaxations of Mayenite, $Y_5Si_3$, and $Ca_2N$ against high-quality experimental data, the accuracy of common density-functional approximations is shown to evolve as interstitial electrons transition from fully confined to increasingly delocalized environments. No single exchange–correlation functional is uniformly optimal across the diverse landscape of electride materials even when considering functionals with a higher placement on the "Jacob's ladder" of sophisticated approaches and computational cost such as HSE06. PBEsol and PBE-D3 in particular systematically over-bound the frameworks and failed to capture the weak, anisotropic confinement of the interstitial electrons, making them unsuitable for quantitative electride modelling.

Remarkably, standard PBE, despite its well-known limitations in describing electron localization and dispersion, performs among the best of the functionals examined here. Across all systems considered, PBE yields lattice parameters in close agreement with experiment, with deviations never exceeding 1% for any lattice vector. This level of accuracy likely reflects fortuitous error cancellation rather than a fully faithful description of electride physics. Nevertheless, this finding carries important implications: much of the existing electride literature based on PBE remains broadly reliable, particularly with respect to qualitative structural trends and initial screening studies. Owing to its numerical robustness and low computational cost, PBE therefore remains a practical and effective tool for large-scale and high-throughput investigations.

The only other functional that also obtained disagreement with experiment less than 1 % is $r^2$SCAN-rVV10, a functional notably cheaper than hybrid approaches such as HSE06. Accordingly, when higher-level treatment is required for quantitative predictions or detailed electronic-structure analysis, r²SCAN-rVV10 represents a more suitable next step beyond PBE than hybrid functionals.

Together, these results support a tiered computational strategy in which PBE can be employed for broad structural exploration and screening, while the more advanced r²SCAN-rVV10 functional can be later are applied selectively. This conclusion is further reinforced by multiple cases in the literature in which PBE-based predictions were later confirmed experimentally, as well as by reports demonstrating qualitative agreement between PBE trends and self-consistent GW calculations.[20,24,34,35,36,23,37,38]

Consideration of the post-processing tools commonly used to define and characterize electrides shows that they must be interpreted with care. Neither ELF nor Bader charge analysis alone provides a complete or universally quantitative descriptor of electride electrons. ELF offers a direct and intuitive visualization of interstitial electron localization and dimensionality, while Bader analysis, when applied to the charge density, constrains how electronic charge is distributed among atomic and interstitial regions but may not always assign separate non-nuclear basins by default. Notably, the recently proposed badELF method explicitly considering charge-density topology, including critical points, and combining this information with ELF, allows for anion electrons to be identified without reliance on a single output. However, human expertise is still ultimately required to draw distinction between electrides and metals. This work therefore supports a balanced and method-agnostic approach to electride identification, in which multiple real-space descriptors are used in concert rather than privileging any single post-processing scheme.

In summary, several practical best practices emerge for the first-principles modelling of electrides. Methods that impose atom-centred corrections, such as DFT+U, should be avoided, as they are ill-suited to describing electrons that reside in interstitial regions. Symmetry operations should be disabled when they are enforced solely on the basis of nuclear positions, as they can artificially constrain the electronic density and obscure off-atom features. Spin polarization should be enabled, since electride electrons may carry magnetic moments that are not associated with atomic sites. For exchange–correlation treatments, standard PBE is generally sufficient for broad structural analysis and initial screening, with more advanced functionals such as r²SCAN-rVV10 applied subsequently for improved quantitative accuracy; by contrast, PBEsol and PBE-D3 consistently over-binds electride systems and should be avoided. Finally, reliable identification of electride behavior requires post-processing analyses that examine both the total charge density and electron localization. Electrides therefore pose both significant challenges and compelling opportunities for the electronic-structure community, demanding careful methodological choices while offering a stringent testbed for advancing first-principles approaches.

## Conflicts of interest

There are no conflicts to declare.

## Data availability

A data availability statement (DAS) is required to be submitted alongside all articles. Please read our full guidance on data availability statements for more details and examples of suitable statements you can use.

## Acknowledgements

The acknowledgements come at the end of an article after the conclusions and before the notes and references.

## Notes and references